# The Emu Sky Knowledge of the Kamilaroi and Euahlayi Peoples

Robert S. Fuller[1,2], Michael G. Anderson[3], Ray P. Norris[1,4], Michelle Trudgett[1]

[1]Warawara - Department of Indigenous Studies, Macquarie University, NSW, 2109, Australia
Email: *robert.fuller1@students.mq.edu.au, michelle.trudgett@mq.edu.au*
[2]Macquarie University Research Centre for Astronomy, Astrophysics and Astrophotonics,
Macquarie University, NSW, 2109, Australia
[3]Euahlayi Law Man, PO Box 55, Goodooga, NSW, 2838, Australia
Email: *ghillar29@gmail.com*
[4]CSIRO Astronomy and Space Science, PO Box 76, Epping, NSW, 1710, Australia
Email: *raypnorris@gmail.com*

## Abstract

This paper presents a detailed study of the knowledge of the Kamilaroi and Euahlayi peoples about the Emu in the Sky. This study was done with ethnographic data that was not previously reported in detail. We surveyed the literature to find that there are widespread reports of an Emu in the Sky across Australian Aboriginal language groups, but little detailed knowledge available in the literature. This paper reports and describes a comprehensive Kamilaroi and Euahlayi knowledge of the Emu in the Sky and its cultural context.

## Notice to Aboriginal and Torres Strait Islander Readers

This paper contains the names of people who have passed away.

## 1. Introduction

Cultural astronomy is the interdisciplinary study of how various cultures have understood and used astronomical phenomena, and the mechanisms by which this understanding is generated (Sinclair, 2006; Iwaniszewski, 2009). It is generally divided into archaeoastronomy (past cultures) and ethnoastronomy (contemporary cultures). Because cultural astronomy is a social science informed by the physical sciences (Ruggles, 2011), the field has been dubbed the "anthropology of astronomy" (Platt, 1991: S76).

Fuller et al. (2014: 3-4) reviewed of the history of cultural astronomy in the Australian Aboriginal context. They report that while there is a rich knowledge of Aboriginal astronomy, the literature on Kamilaroi and Euahlayi astronomy, based on ethnography from the 19th century, was often very limited in detail and contained many contradictions between the stories reported. For that reason, this project (consisting of Robert Fuller, Ray Norris, and Michelle Trudgett) included an ethnographic phase to collect knowledge of the sky from current Kamilaroi, Euahlayi, and neighbouring communities. The ethnography comprised multiple interviews and recordings of stories during 2013 from eight participants with mostly mixed heritage from the Kamilaroi, Euahlayi, Ngemba, and Murrawarri communities. Those participants are noted in the Acknowledgements. One participant, Michael Anderson, with both Euahlayi and Kamilaroi heritage, provided such a complete description of the "Emu in the Sky" as it related to his culture that he has been included as an author of this paper. The "Kamilaroi Project" (as we will continue to describe the study conducted by Fuller, Norris and Trudgett) confirmed the hypothesis that the knowledge gained could add to the current body of knowledge of Australian Aboriginal sky culture. Most of the data was released under the terms of the Ethics Approval by Macquarie University. This paper presents previously unpublished data from the Kamilaroi Project used to determine whether the knowledge about the Emu in the Sky collected through the larger project adds a deeper level of understanding into the sky culture of the Kamilaroi and the Euahlayi peoples.

Like most Aboriginal stories, those collected in the Kamilaroi Project do not just entertain and describe some physical object in the sky. Aboriginal culture is oral in nature, and oral transmission of knowledge is extremely important, particularly in regards to Law. Aboriginal Law governs all aspects of Aboriginal life, establishing a person's rights and responsibilities to others, the land, and natural resources (Law Reform Commission of Western Australia, 2006: 64). Cultural stories transmit Law, and in this respect can have different levels of meaning. Sveiby and Skuthorpe (2006: 45-51) describe four levels: one being for children (to explain nature), others being for relationships between people, relationships between the community and country, and ceremonial practices. A participant in the Kamilaroi project said that some stories could have up to "30 levels" of meaning, suggesting that most of those levels were secretive and ceremonial in nature. Here we avoid references to secret levels.

## 2. The Kamilaroi and Euahlayi Peoples

The Kamilaroi and Euahlayi peoples are an Australian Aboriginal cultural grouping located in the north and northwest of New South Wales (NSW). The Kamilaroi language groups are described as "Gamilaraay" and "Yuwaalaraay/Yuwaalayaay" (Ash, et al., 2003: 1), while the Euahlayi have a similar but distinct language.

The geographical boundaries of the area defined for this study of the Kamilaroi and the Euahlayi are based on the language group boundaries reported by Austin (2008: 2), which is very similar to that proposed by Sveiby and Skuthorpe (2006: 25). Figure 1 shows the approximate area of the cultural group and languages. Participants in the Kamilaroi project have stated that the map is incorrect, in that Yuwaalayaay and Yuwaalaraay are gradations of the same Kamilaroi clan dialect, that this clan grouping, indicated by the area "6", is actually further north, and that the Euahlayi language group covers the area "7", the southern part of area "6", and westwards towards the Culgoa River. Discussion with several linguists has confirmed that a debate continues about the locations and names of some of the language groups in this area.

The population of this cultural grouping was estimated at 15,000 in 1788, and as low as 1,000 in 1842 (Sveiby and Skuthorpe, 2006: 25-26). A participant in the Kamilaroi project has speculated this could have been as large as 60,000 with the resources available in the area of the study. As a result of pressure from European settlers, there was a displacement of Aboriginal people in this group towards the northwest. The current population of people identifying as Kamilaroi and/or Euahlayi ancestry is approximately 29,000 (Kamilaroi Nation Applicant Board, 2013. pers. comm. Board Chairman).

## 3. The Emu in the Sky Across Australia

We searched for references to an Emu in the Sky (Emu) across Australian Aboriginal literature, and found them in the literature from cultures in South Australia (SA) (Nullarbor and central desert), West Australia (WA) (Kimberley, Tanami, and Murchison regions), Northern Territory (NT), Victoria (VIC), Queensland (QLD) (Gulf country and southeast), and NSW (Sydney basin).

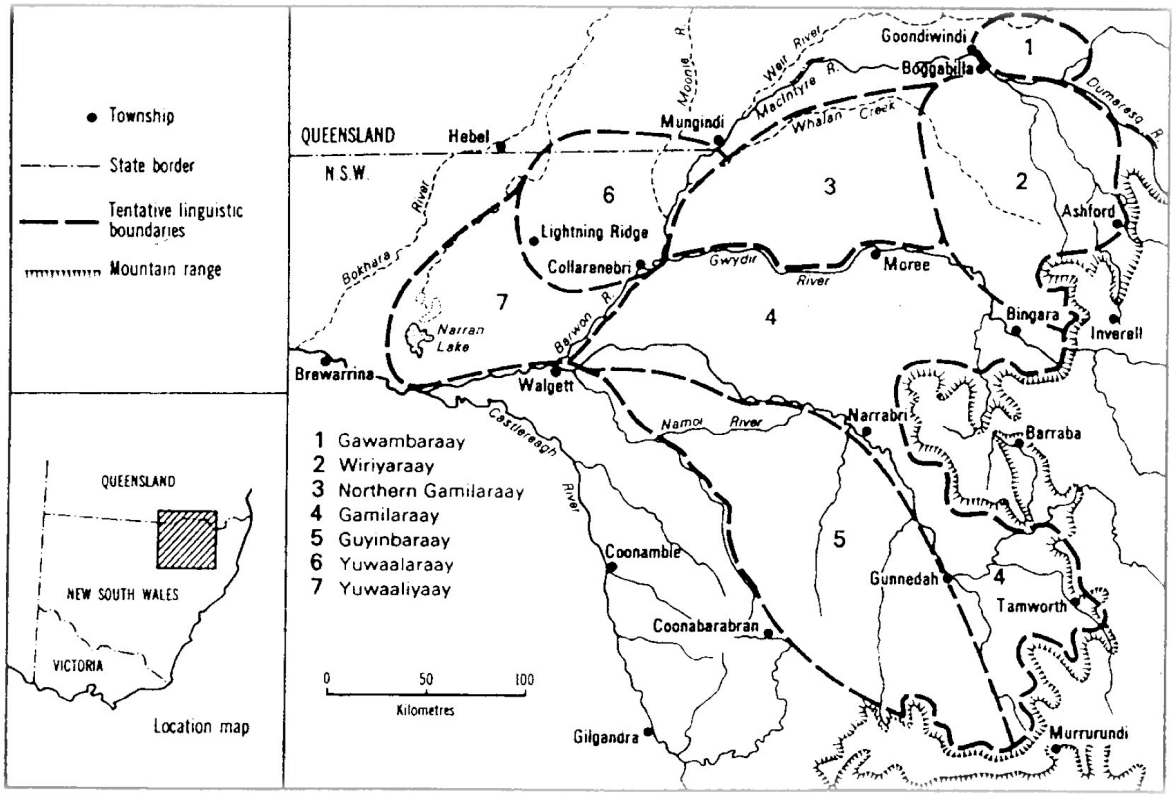

*Figure 1: Language location in northern New South Wales (Austin, 2008) note: 1-7 are Kamilaroi dialects*

The earliest reference to the Emu was by Stanbridge (1857: 139), who reported that the Boorong people of west central VIC said that an emu (Tchingal) resided in the dark patch (the Coalsack nebula) under the constellation Crux (commonly known as the Southern Cross). The next mention in the literature was by Ridley (1873: 273-274), who spent an evening under the sky with an Aboriginal man from near Walgett, NSW called King Rory, who informed Ridley that there was an Emu (*gao-ergi*) in "the dark space under the tree", meaning the Coalsack (the tree being the Southern Cross). King Rory, who was most likely Euahlayi, used the term *gao-ergi*, which is phonetically very similar to the current Kamilaroi/ Euahlayi words *Gawarrgay/ Gawarghoo* (Ash, et al., 2003: 82). We have confirmed that *Gawarrgay/ Gawarghoo* is the correct word for the Emu, as *dhinawan* is used for the emu on Earth. Fuller et al. (2014: 29) argued that King Rory had been taught this knowledge by his grandfather when he was around 15, and Ridley estimated him to be about 60 in 1871, so his description of the Emu was learned well before any European explorers or settlers had reached the Walgett area, giving strong support to the idea that the Emu is pre-European contact in origin. Ridley also gave King Rory's tribal name as *Ippai Dinoun*, *Ippai* being one of the Euahlayi marriage classes, and *Dinoun* being Ridley's spelling of the current *dhinawan*, which is the emu's name in Kamilaroi/Euahlayi, and was his totem, so he should have been knowledgeable about the Emu.

Later references include Palmer (1885: 174) who has a reference from the Gulf country in QLD, Bates and Wilson (1972: 59-60), and Bates (n.d.: 13) who said that the emu in the "Yamminga times of long ago," went up into the sky and "became Wej Mor – the dark patch in the Milky Way". Bates is believed to have collected this story at Ooldea, SA, around 1904 from the Ngalea language group (Great Victoria Desert).

Basedow (1925: 315, 332-334) has several references to the Emu from northern Australia. Some unknown Aboriginal groups from the Musgrave Ranges of the Tanami Desert (WA) spoke of a "resting emu" (*kaleya pubanye)* in the Coalsack. The Larrakia from near Darwin (NT) had a very complete view of the Emu which is remarkably like that of more recent investigations, and told Basedow that "the Coalsack was the head of a gigantic emu" which was made up of dark patches in the Milky Way as far as Scorpius, with the legs extending further.

In his field notes of an expedition to the Warburton Ranges, Tindale (1935: 457-459) refers to a story from the Pitjantjatjarra (Central Desert) about an Emu called *Kalaia*. Worms (1940: 271) has a reference to the Emu from his work with a Kimberley (WA) group.

Love (1987: 4) refers to his correspondence with V. Ford in 1985, in which Ford describes the Emu as follows: "to the Aboriginal this dark constellation was the Emu, its head being the Coalsack, its body being in Scorpius and its legs in Ophiuchus". In Hafner et al. (1995: 34) Ngitji Ngitji told of stories of the Emu and the Milky Way from northern SA. More recently, Cairns (1996: 9-10) suggested that a rock engraving of an emu at the Elvina Track site in Kuringai National Park (NSW) could represent the Emu. Norris and Norris (2009: 6-7) have shown that the engraving mirrors the Emu, in both shape and azimuth, in April, which is the time of the year when emus lay their eggs (Figure 2).

Cairns and Harney (2003) describe the "Cosmic Emu" of the Wardaman people and their neighbours in the area bounded by the Victoria and Daly Rivers of the NT, based on the knowledge of Bill Yidumduma Harney, a Wardaman elder. They connect the Emu to songlines and rock art, and to descriptions by previous writers.

## 4. The Emu of the Kamilaroi and Euahlayi

There are a limited number of written sources about the culture of the Kamilaroi and Euahlayi, mostly from the latter half of the 18th century. These were Ridley (1856, 1873, 1875), Fraser (1888), Greenway (1878, 1901), and Fison and Howitt (1880). R.H. Mathews (1900 and 1904) are relevant to this study, as are papers by K. Langloh Parker, a contemporary of Mathews who lived on the Narran River in the late 1800's and collected a large body of folklore about the Euahlayi (Parker, 1896, 1898, 1914; Parker and Lang, 1905). Sveiby and Skuthorpe (2006) have more recently described the culture of the Nhunggabarra band of the Kamilaroi/Euahlayi language group.

It has been established that the idea of an Emu, as a cultural object used in stories, rather than the emu bird, existed across Australia at the time of European invasion. Basedow's information from the Larrakia, Ford's description, and the more recent investigation of the Kuringai people's stone engraving of an emu near Sydney (Figure 2), all point to at least some Aboriginal groups seeing the Emu as a long, stretched out figure in the dust clouds of the Milky Way from the Coalsack to beyond Scorpius. Six of the eight participants in the ethnographic phase of the Kamilaroi project referred to the Emu as an emu figure stretching from the Coalsack to Scorpius, and most could describe its appearance. However, the story from Anderson, as used in this study, is the only complete story collected, including linkage to the Kamilaroi/Euahlayi culture and resource management.

The Emu, as seen by the Kamilaroi and Euahlayi, changed in position from season to season, as the Milky Way containing the Emu changed position in the night sky. As the Emu changes position, it alters in appearance, and that appearance has connections to cultural and resource matters. All of the images of the Emu's appearance in this study are seen mid-evening, around 21:00 local time, in the area of the study. The head of the Emu (the Coalsack) can be seen all year, as the Southern Cross is circumpolar when viewed from this latitude, but the neck and the body of the Emu may or may not be visible at different times of the year, due to the rotation of the sky (which is the result of the tilt of the Earth from summer to winter, and the position of the Earth in its rotation around the Sun). We have used the computer planetarium program, Starry Night Pro[1], adjusted to provide a non-light polluted sky to determine the various views of the Emu used in this study. Anderson describes the cultural appearance of the Emu in the Milky Way as seen using Starry Night Pro. Through this, the authors were able to confirm that the appearance of the Emu began at the Coalsack under the star α Crucis, which formed the Emu's head, then β and α Centauri, which form the start of the neck, down the dust lanes of the Milky Way to η Lupus and $γ^2$ Norma, at which point the dust lanes expand with the body of the Emu, reaching the maximum thickness with ε and λ Scorpii, and tapering towards 36-Ophiuchi and 3-Sagittarii, eventually ending near μ Sagittarii.

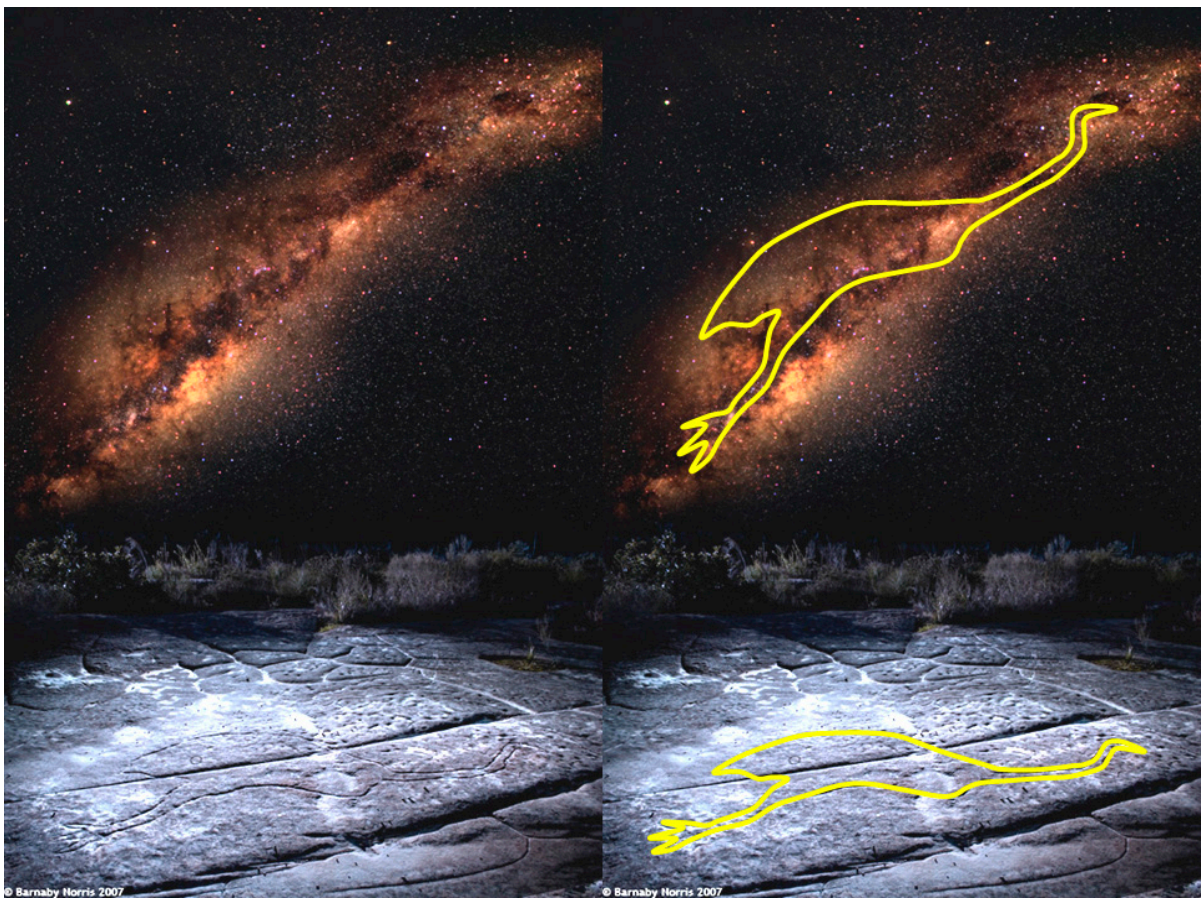

*Figure 2: Kuringai Emu in the Sky. Images courtesy Barnaby Norris and Ray Norris.*

While the head and neck of the Emu can be seen in the sky as early as March, it reaches its first appearance in full length after sunset in April and May, when it is seen stretching from the South to the southeast (Figure 3). At this time, the Kamilaroi and Euahlayi say the Emu has legs, and appears to be running. This reflects the behaviour of female emus, who chase the males during the mating season[1]. Because emus are laying their eggs at this time, the appearance of the celestial Emu is a strong reminder to the Kamilaroi and Euahlayi people that eggs are available.

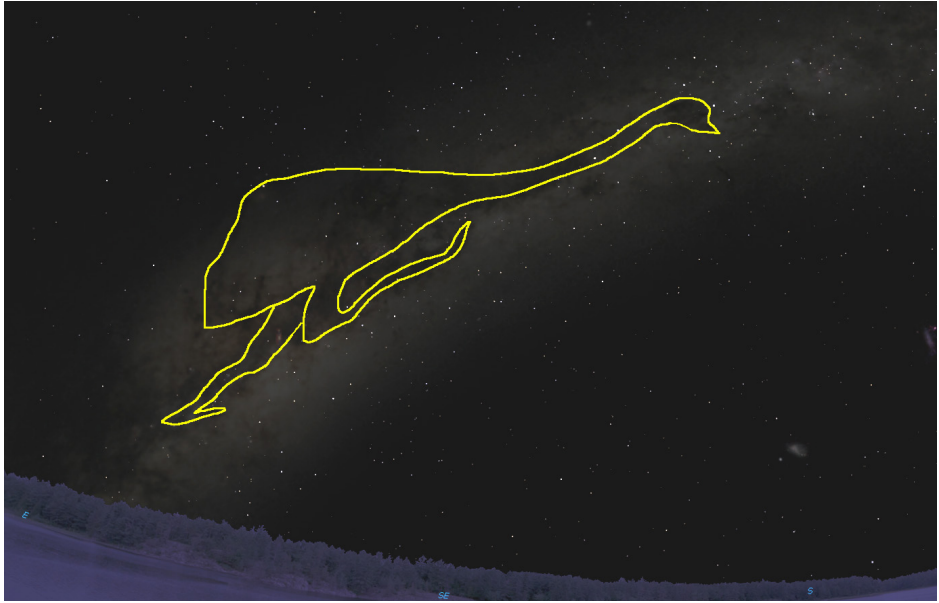

*Figure 3: The Emu in autumn (April-May), running after a mate. Image from Starry Night Education.*

In June and July, the appearance of the Emu changes. The legs disappear, and the Emu, which is now seen as male, is sitting on its nest, hatching the new chicks (Figure 4). After female emus lay the eggs, it is the males that broods the eggs (*ibid*). At this time, the eggs are still an available resource, and can be taken from the nest.

The Kamilaroi and Euahlayi have in common their male initiation ceremony, the *bora*. Many language groups in southeast Australia used a similar ceremony, sometimes using the borrowed Kamilaroi word, *bora*. For the Kamilaroi and Euahlayi, the preferred months for their *boras* are after August, according to participants in the Kamilaroi project, because the Milky Way was vertical in the sky to the southwest in August and early September.

The connection between the Milky Way and the *bora* could be linked to the culture hero, *Baiame*, who was common to many language groups in southeast Australia (Fraser, 1888; 10). *Baiame's* son, *Daramulan,* was given to the people and it is through *Daramulan* that *Baiame* "sees all" (Fraser 1882: 208, Howitt 1884: 458). *Baiame* is worshipped at the *bora* ceremony (Ridley, 1873: 269) and *Daramulan* is believed to come back to the Earth by a pathway from the sky (Fraser, 1882: 212). Eliade and Sheed (1996: 41) reports that *Baiame* "dwells in the sky, beside a great stream of water (Milky Way)".

In late winter (August to September), the neck of the Emu becomes indistinct in the sky, leaving the body to represent an emu egg (Figure 6). This was taken, according to Anderson, as a sign that the emu chicks were hatching, and that the egg resource was no longer available. Because the male emus look after the chicks (Eastman, 1969), this has led to some speculation (Love, 1987: 3) that this connects the Emu to the *bora* ceremony. As the male emu hatches and raises the chicks, so the Aboriginal elders nurture the male initiates. There is little literature on this subject, other than Winterbotham (1957: 3-4), who reported on information from a southeast QLD Aboriginal elder, Gaiarbau, who indicated the link between the *bora* and sky. This reinforces Love's speculation.

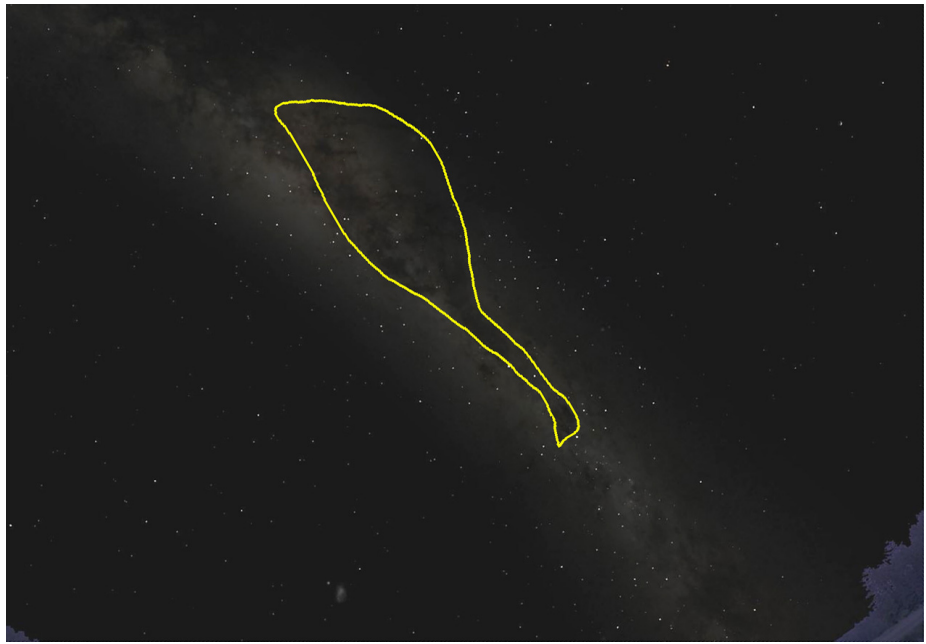

*Figure 4: The Emu in winter (June-July), sitting on its nest (image courtesy of Starry Night Education)*

Fuller et al. (2013: 36) tested the hypothesis that the orientations of *bora* ceremonial circles are aligned to the position of the Milky Way, which is vertical in the sky after sunset to the south-southwest in August and early September. Figure 5 shows the preference in orientation of 68 *bora* sites in NSW and southeast QLD where the orientation was known. The results show that the preference is strong in the southern quadrant, where the Milky Way is located in August and early September. To determine if this was a chance clumping of a random distribution of orientations, Fuller et al. conducted a Monte Carlo simulation, in which 68 orientations were distributed randomly in each of the bins shown in Figure 5. They repeated this process 100 million times. In only 303 of the 100 million runs did the number in any one bin equal or exceed 35, from which they concluded that the likelihood of the peak in Figure 5 occurring by chance is about $3 \times 10^{-6}$ or 0.0003%. They concluded that this distribution was clearly not the result of chance, and that the builders of the *bora* rings intentionally aligned most of them to the southern quadrant. This lends support to the claims that bora ceremonies are linked to the Milky Way.

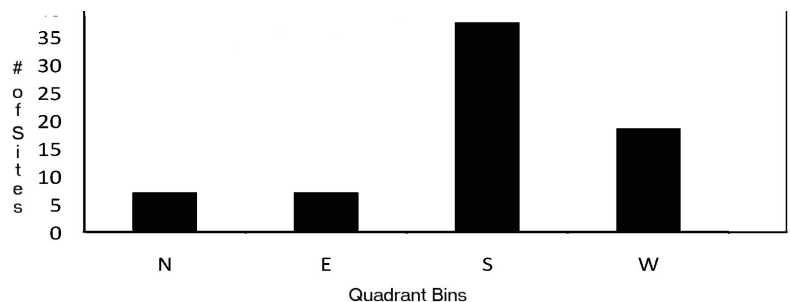

*Figure 5: Frequency of alignment of bora sites to cardinal quadrants (defined as a quarter of a circle; an arc of 90°, N being 0°, S being 180°), c.f. Fuller et al. (2013).*

The head of the Emu is still visible in the sky in late winter, and together with the body, they appear to form a large and small ring in the Milky Way, which may be representative of the small and large *bora* rings that are laid out on the ground (Figure 6). The head represents the smaller, sacred, *bora* ring, and body the larger, public ring, and looking at the rings in the sky, they mirror the layout of the *bora* rings on the ground. If Fuller et al. (2013) are correct in their alignment hypothesis, this may be the reason that *bora* sites are aligned to the southern quadrant. At this time of year, Aboriginal people in the area of the study would be leaving their winter camps to travel to ceremonial sites for ceremonies including the *bora*.

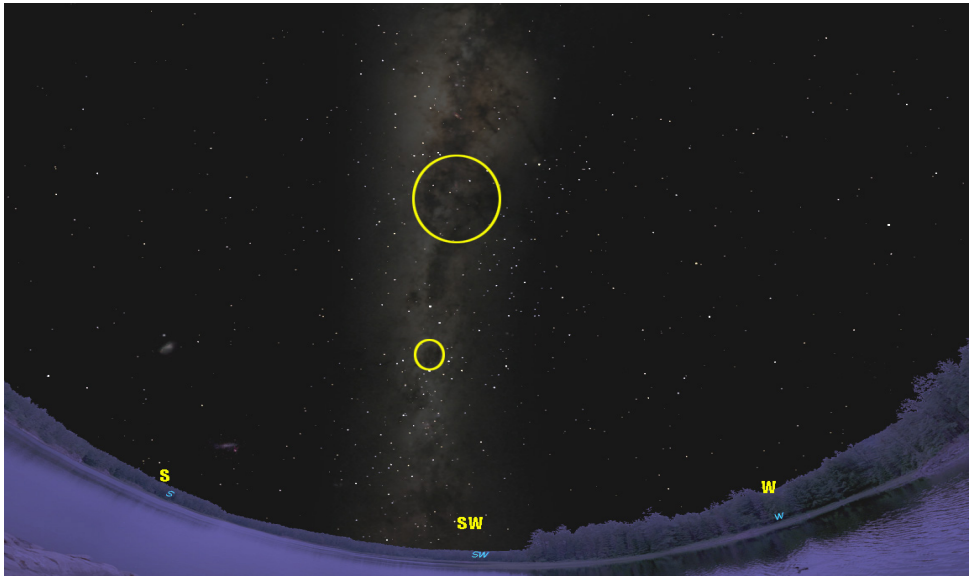

*Figure 6: Emu in late winter (August-September) over the South-southwest. Image from Starry Night Education.*

In the spring, around November, the Emu once again is transformed (Figure 7). For the Kamilaroi and Euahlayi, the Emu is also *Gawarrgay/Gawarghoo*, a featherless emu who travels to waterholes and looks after everything that lives there. Come November, the Emu is now (along with the Milky Way it inhabits) low on the horizon in the evening. Due to atmospheric extinction, the neck and the head are difficult to see, so the body of the Emu seems to be "sitting" on the horizon. According to Anderson, this is because the Emu is now believed to be sitting in a waterhole, and as a consequence, the waterholes in country are believed to be full (which would normally be the case in southeast Australia after the winter rains). The Kamilaroi have another name for the emu bird: *ngurran.gali* (*Euahlayi: dthnarwon.gulli*). This translates to "an emu sitting" or "emu in the water", which may well relate to this view of the celestial Emu.

Later in the summer, the Milky Way and the Emu have dipped even lower, and the Emu has become almost invisible on the horizon (Figure 8). At this time, the Emu is believed to have left the waterholes, and because of this, the waterholes in country are dry, which may well be the case in the summer. The Emu will not be visible again until its head peeks above the horizon in February, followed by the body in March.

Some of the major themes of the Kamilaroi Project, such as "what's up there is down here", are reflected in the Kamilaroi/Euahlayi stories of the Emu. A few participants, including

Anderson, commented on their belief that, at one time, the sky and everything in it was "down here", and what is now "down here" was in the sky. For that reason, what is seen in the sky now is also on the ground, and the varying views of the Emu also have close connections with things on the ground, in particular the emu bird, which was an important resource. The view of the Emu was closely connected to the resource management of the emu, possibly the ceremonial aspects of the male initiation ceremony, and in regards to waterholes, the management of country.

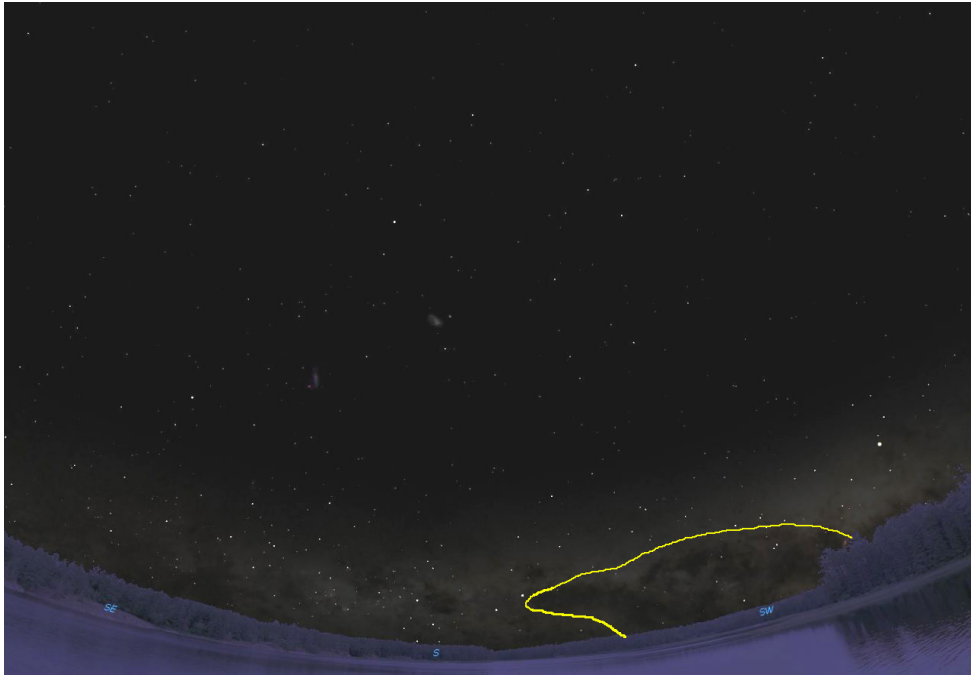

*Figure 7: The Emu sitting in a waterhole in spring (November). Image from Starry Night Education.*

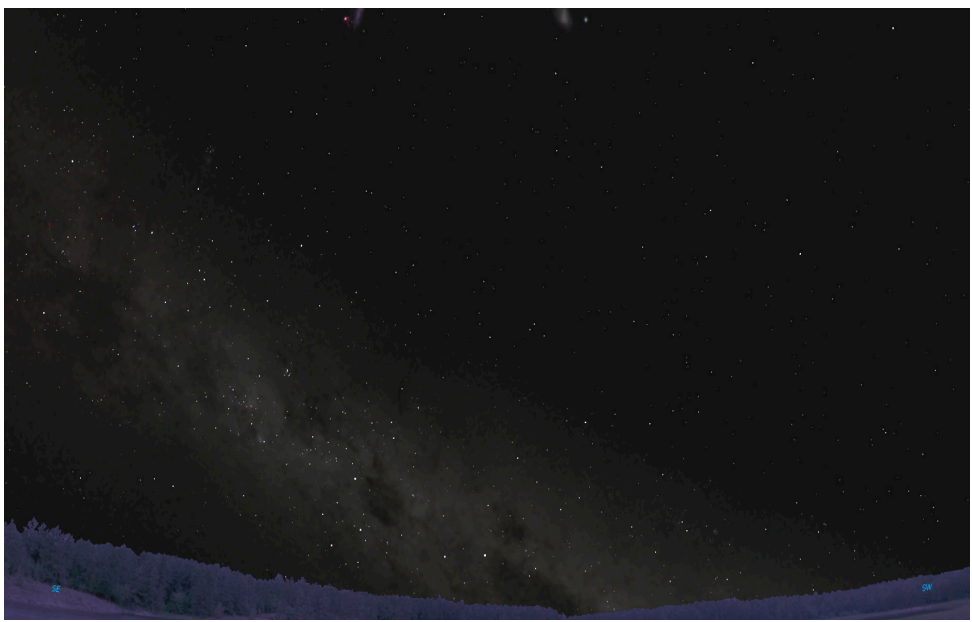

*Figure 8: Summer, when most of the emu is below the horizon. Image from Starry Night Education.*

## 5. Conclusions

We show that the Emu in the Sky is an important cultural figure in many different parts of Aboriginal Australia. While many of the reports in literature have only the barest details, where traditional knowledge still exists, there is the possibility of working with the knowledge holders to restore a more complete understanding to the public record. The support of the participants in the Kamilaroi Project has provided a very detailed picture of how the Emu in the Sky fitted into the sky knowledge and culture of the Kamilaroi, Euahlayi, and possibly their Murrawarri and Ngemba neighbours. This would appear to be the first Aboriginal cultural grouping in Australia where this knowledge was brought together into a unified description. We believe the knowledge gained in this one aspect of culture meets the aims of the hypothesis of the Kamilaroi Project to add to the overall understanding of the sky knowledge of the Kamilaroi and Euahlayi peoples. This information may also be used in future research to strengthen the understanding of how Australian Aboriginal culture linked the behaviour of cultural objects in the sky with animals and resources on the land.

## Acknowledgements

We acknowledge and pay our respects to the traditional owners and elders, both past and present, of the Kamilaroi, Euahlayi, Ngemba, and Murrawarri peoples. We thank the participants of the Kamilaroi project, Michael Anderson, Rhonda Ashby, Lachie Dennis, Paul Gordon, Greg Griffiths, Brenda McBride, Jason Wilson, and one anonymous person, for their stories and those of their families. We particularly thank Michael Anderson for his detailed story on the Emu, which is central to this study. This paper was originally presented at the 2013 Australian Space Sciences Conference at the University of New South Wales. We also thank the two anonymous referees and the guest editor, Duane Hamacher, for their feedback.

The ethnographic data from the Kamilaroi Project will be retained by Robert Fuller for five years, and then transferred for secure storage by Macquarie University in accordance with the terms of the Ethics Approval. Access to the data, subject to the terms of the Ethics Approval, can be requested through the author.

## Notes

1. http://astronomy.starrynight.com

## References

Ash, A., Giacon, J., and Lissarrague, A., 2003. *Gamilaraay, Yuwaalaraay and Yuwaalayaay Dictionary*. Alice Springs, IAD Press.

Austin, P., 2008. The Gamilaraay (Kamilaroi) language, northern New South Wales - a brief history of research. In McGregor, W. (ed). *Encountering Aboriginal languages: studies in the history of Australian linguistics*. Canberra, Pacific Linguistics. Pp. 37-57.

Basedow, H., 1925. *The Australian Aboriginal*. Brooklyn, NY, AMS Press.

Bates, D. and Wilson, B., 1972. *Tales Told to Kabbarli: Aboriginal Legends Collected by Daisy Bates.* Sydney, Angus and Robertson.

Bates, D., n.d. *Papers of Daisy Bates*. Archives of the South Australian Museum, Adelaide. http://archives.samuseum.sa.gov.au/aa23/serieslist.htm

Cairns, H., 1996. *Sydney Rock Carvings and Astronomy*. Unpublished Manuscript, Hugh Cairns.

Cairns, H. and Harney, B., 2003. *Dark Sparklers*. Sydney, Hugh Cairns.

Eastman, M., 1969. *The life of the emu*. Sydney, Angus and Robertson.

Eliade, M. and Sheed, R., 1996. *Patterns in Comparative Religion*. Lincoln, University of Nebraska Press.

Fison, L. and Howitt, A., 1880. *Kamilaroi and Kurnai.* Sydney, George Robertson.

Fraser, J., 1882. The Aborigines of New South Wales. *Transactions and Proceedings of the Royal Society of New South Wales*, 26, 193-233.

Fraser, J., 1888. *The Aborigines of Australia: their Ethnic Position and Relations*. London, The Victoria (Philosophical) Institute.

Fuller, R.S., Hamacher, D.W., and Norris, R.P., 2013. Astronomical Orientations of Bora Ceremonial Grounds in Southeast Australia. *Australian Archaeology*, 77, 30-37.

Fuller, R.S., Norris, R.P., and Trudgett, M., 2014. *The Astronomy of the Kamilaroi People and Their Neighbours*, submitted to the journal *Australian Aboriginal Studies* on 1 November 2013. Preprint: http://arXiv.org/abs/1311.0076

Greenway, C., Honery, T., McDonald, Rowley, J., Malone, J., and Creed, 1878. Australian Language and Traditions. *The Journal of the Anthropological Institute of Great Britain and Ireland,* 7, 232-74.

Greenway, C., 1901. Berryberry, Aboriginal Myth. *Science of Man*, 4(10), 168.

Hafner, D., Ngitji Ngitji, M. and Subasinghe, S., 1995. Traditional storytellers. *Australian Feminist Studies*, 10(21), 33-38.

Howitt, A., 1884. On Some Australian Ceremonies of Initiation. *The Journal of the Anthropological Institute of Great Britain and Ireland*, 13, 432-459.

Iwaniszewski, S., 2009. Por una astromiá cultural renovada. *Complutum*, 20, 23-37.

Law Reform Commission of Western Australia, 2006. *Australian Customary Law, Final Report, Project 94*. Perth, Law Reform Commission of Western Australia. URL: http://www.lrc.justice.wa.gov.au/P/project_94.aspx

Love, W., 1987. There is an emu on the bora ground. *Anthropological Society of Queensland Newsletter*, 177, 3-4.

Mathews, R., 1900. The Origin, Organization and Ceremonies of the Australian Aborigines. *Proceedings of the American Philosophical Society, 39(164)*, 556-78.

Mathews, R., 1904. Ethnological Notes on the Aboriginal tribes of NSW. *Proceedings of the Royal Society of New South Wales,* 10, 203-381.

Norris, R.P., and Norris, P.M., 2009. *Emu Dreaming: An Introduction to Aboriginal Astronomy*. Sydney, Emu Dreaming.

Palmer, E., 1885. Concerning Some Superstitions of North Queensland Aborigines. *The Proceedings of the Royal Society of Queensland*, 2(2), 1-13.

Parker, K., 1896. *Australian Legendary Tales*. London, D. Nutt.

Parker, K., 1898. *More Australian Legendary Tales*. London, D. Nutt.

Parker, K., 1914. A Kamilaroi Legend of the Southern Cross. Adelaide, *The Mail, Saturday, 11 April*, 8.

Parker, K. and Lang, A., 1905. *The Euahlayi Tribe, a Study of Aboriginal Life in Australia*. London, A. Constable.

Platt, T., 1991. The Anthropology of Astronomy. *Archaeoastronomy*, 16, S76-S81.

Ridley, W., 1856. Kamilaroi Tribe of Australians and Their Dialect, in a Letter to Dr Hodgkin. *Journal of the Ethnological Institute of London*, 4, 285-293.

Ridley, W., 1873. Report on Australian Languages and Traditions. *The Journal of the Anthropological Institute of Great Britain and Ireland*, 2, 258-291.

Ridley, W., 1875. *Kamilaroi, and other Australian Languages.* NSW, Thomas Richards, Government Printer

Ruggles, C.L.N., 2011. *Pushing back the frontiers or still running around in circles? Interpretive archaeoastronomy thirty years on*. In Ruggles, C.L.N. (Ed.). *Archaeoastronomy & Ethnoastronomy: Building Bridges Between Cultures*. Cambridge, Cambridge University Press. Pp. 1-18.

Sinclair, R.M., 2006. *The Nature of Archaeoastronomy*. In Bostwick, T.W. and Bates, B. (eds) *Viewing the Sky Through Past and Present Cultures; Selected Papers from the Oxford VII International Conference on Archaeoastronomy*. Phoenix, AZ, Pueblo Grande Museum Anthropological Papers 15. City of Phoenix Parks and Recreation Department. Pp. 13–26.

Stanbridge, W.E., 1857. On the Astronomy and Mythology of the Aborigines of Victoria. *Proceedings of the Philosophical Institute of Victoria, Transactions*, 2, 137-140.

Sveiby, K. and Skuthorpe, T., 2006. *Treading Lightly: The Hidden Wisdom of the World's Oldest People*. Sydney, Allen & Unwin.

Tindale, N., 1935. *Unpublished fieldwork notes*. South Australia Museum, Adelaide. Catalogue Number aa318/1/14. Pp. 457-459.

Winterbotham, L.P., 1957. *Gaiarbau's story of the Jinibara tribe of southeast Queensland (and its neighbours)*. Manuscript (MS 45) held at the Australian Institute for Aboriginal and Torres Strait Islander Studies, Canberra.

Worms, E., 1940. Religious Ideas and Culture of Some North-Western Australian tribes in fifty legends. *Annali Lateranensi 2*. Rome, Pontifico Museo Missionario Etnologico.

**About the Authors**

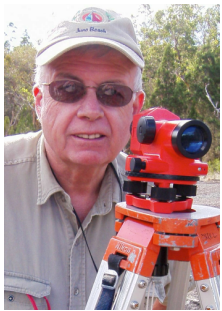

Robert (Bob) Fuller is a late-life returnee to academia. He was educated as an anthropologist (B.A., Gettysburg College, USA), but never practiced as one. After retiring from careers as a military pilot then businessman, with a long interest in astronomy, he took up the challenge to combine his interests in anthropology and astronomy by studying Australian cultural astronomy. Mr Fuller is now studying Kamilaroi astronomy for a Master of Philosophy in Indigenous Studies at Macquarie University.

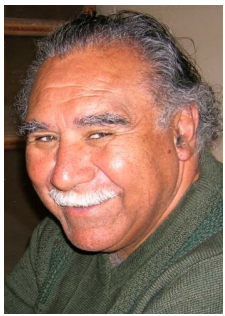

Michael Anderson (Nyoongar Ghurradjong Murri Ghillar) is an Aboriginal rights activist, leader of the Euahlayi tribe of 3,000 people living in northwestern New South Wales, and Native Title claimant to their traditional lands on their behalf. He was taught Euahlayi customs and traditions through his people's sacred ceremonies. Mr Anderson has lectured in Aboriginal studies and Aboriginal politics at several Australian universities, where he wrote and taught units in Aboriginal studies that were inclusive of traditional Aboriginal society.

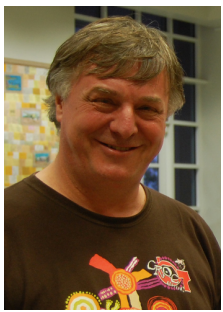

Professor Ray Norris is an astrophysicist with CSIRO Astronomy & Space Science and an Adjunct Professor at Warawara - Department of Indigenous Studies at Macquarie University. He has published about 300 peer-reviewed papers, including 15 on Aboriginal Astronomy, and wrote the book "*Emu Dreaming: an Introduction to Australian Aboriginal Astronomy*".

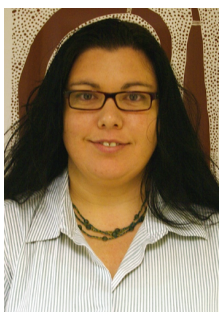

Associate Professor Michelle Trudgett is an Indigenous scholar from the Wiradjuri Nation in New South Wales. She is the Head of Warawara - Department of Indigenous Studies at Macquarie University. Her current ARC funded research seeks to create a model of best practice for the supervision of Indigenous doctoral students. A/Prof Trudgett is passionate about developing strategies to ensure Indigenous higher education students receive culturally appropriate support throughout their academic journeys.